\documentstyle[aps,floats,rotate,multicol]{revtex}

\input epsf         
\epsfverbosetrue    

\begin{document}

\twocolumn[\hsize\textwidth\columnwidth\hsize\csname@twocolumnfalse%
\endcsname

\title{Interlayer pair tunneling and gap anisotropy in YBa$_2$Cu$_3$O$_{7-\delta}$}
\author{Wen-Chin Wu}
\address{Department of Physics, Simon Fraser University, Burnaby,
British Columbia, Canada V5A 1S6}
\address{Department of Physics, National Taiwan Normal University, Taipei 11650,
Taiwan\cite{wu}}
\date{\today}
\maketitle
\draft

\begin{abstract}
Recent ARPES measurement observed
a large $ab$-axis gap anisotropy, $\Delta(0,\pi)/\Delta(\pi,0)=1.5$,
in clean YBa$_2$Cu$_3$O$_{7-\delta}$.
This indicates that some sub-dominant component may exist in
the $d_{x^2-y^2}$-wave dominant gap.
We propose that the interlayer pairing tunneling contribution can
be determined through the investigation of the order parameter anisotropy.
Their potentially observable features in
transport and spin dynamics are also studied.
\end{abstract}
]

The symmetry of the order parameter continues to be a fundamental
issue in the studies of high-$T_c$ cuprate superconductors.
While the order parameter of cuprates
is extensively believed to be dominated by
$d_{x^2-y^2}$-wave, whether there exists a sub-dominant component
remains an open question \cite{kouznetsov97}.
Recent angle-resolved photoemission spectroscopy (ARPES) measurement
\cite{lu01} has revealed a
significant superconducting
gap anisotropy in clean untwinned YBa$_2$Cu$_3$O$_{7-\delta}$ (YBCO).
The excitation gap along the $k_y$ axis is found to be 50\%
more than that along the $k_x$ axis. This strongly suggests in YBCO that
some sub-dominant component is involved in the $d_{x^2-y^2}$-wave dominant
order parameter. It is important to ask
what the symmetry and the mechanism of the sub-dominant component are.

In terms of structure and electronic properties,
YBCO differs from La$_{2-x}$Sr$_x$CuO$_{4+\delta}$ (LSCO) or
Bi$_2$Sr$_2$CaCu$_2$O$_{8+\delta}$ (BSCCO) in one crucial aspect.
Due to the existence of one-dimensional (1D) CuO chain, the system
is orthorhombic rather than tetragonal. The perfect $D_{4h}$ symmetry
is broken which accounts naturally for the
additional symmetry mixed in the order parameter.
In this case, the most favorable order parameter is
the so-called $d$+$s$ model, $\Delta_{\bf k} \sim \cos (2\phi)+s$
[$\phi$ is the azimuthal angle of ${\bf k}$ on the Fermi surface (FS)],
which arises providing that the
in-plane pairing interaction is given by a separable
$g({\bf k,k^\prime})=-V f({\bf k})f({\bf k}^\prime)$ with $V>0$ and
$f({\bf k})=\cos (2\phi)+s$. The nodal lines of this gap
are deviated from the usual diagonals ($k_x=\pm k_y$) and
consequently the gap magnitude is anisotropic between the
$k_x$ and $k_y$ axes.
The ARPES data, $\Delta(0,\pi)/\Delta(\pi,0)=1.5$,
would correspond to $s=-0.2$ in this simple $d$+$s$ model.

Another complication occurs in YBCO (and also in BSCCO) because it involves
more than one conducting layer within a unit cell. This enables
the interlayer
Cooper pair tunneling in the superconducting state.
Following Chakravarty {\em et al.} \cite{CSAS93},
a superconducting {\em bilayer} is considered to which
an {\em interlayer} pair tunneling Hamiltonian

\begin{eqnarray}
H_\perp=\sum_{\bf k}T_{J}({\bf k})
a^{\dagger}_{1{\bf k}\uparrow}a^{\dagger}_{1{\bf -k}\downarrow}
a_{2{\bf -k}\downarrow}a_{2{\bf k}\uparrow}
+{\rm H.c.}
\label{eq:model}
\end{eqnarray}
coexists with the usual {\em intralayer} pairing interaction term.
Here the index 1 (2) denotes for layer 1 (2) and
$T_J({\bf k})$ is the interlayer Josephson pair tunneling coupling.
The momentum-conserved interlayer pair tunneling term
(\ref{eq:model}) can be generated from the interlayer single-particle hopping
$\sim\sum_{\bf k}t_\perp({\bf k})a^{\dagger}_{1{\bf k}}a_{2{\bf k}\uparrow}$,
where the hopping integral
$t_\perp({\bf k})$ can be determined from the band structure
\cite{anderson94,xiang96}.
Thus $T_J({\bf k})=t_\perp({\bf k})^2/t$ is usually assumed with
$t$ the nearest-neighbor hopping.
As pointed out by Chakravarty and Anderson \cite{CA94},
due to the highly non-Fermi liquid nature of cuprates, the coherent interlayer
single-particle hopping is blocked, which nevertheless
gives rise to the interlayer pair tunneling.

Combining (\ref{eq:model}) and the intralayer pairing term,
one can write down the BCS gap equation for layer $i$ (1 or 2)

\begin{eqnarray}
\Delta_{{\bf k}i}=-\sum_{\bf k^\prime}g({\bf k,k^\prime})
\langle a_{i{\bf -k^\prime}\downarrow}a_{i{\bf k^\prime}\uparrow}\rangle
+T_J({\bf k}) \langle a_{j{\bf -k}\downarrow}a_{j{\bf k}\uparrow}\rangle,
\label{eq:gap}
\end{eqnarray}
where $j\neq i$ and the angular bracket denotes the anomalous mean-field average.
It should be stressed that
the two (CuO$_2$ plane) layers considered here for YBCO
are mediated and renormalized by
the third CuO chain layer and appear to be {\em orthorhombic} instead of tetragonal.
The issue is then how the interlayer pair tunneling plays the role between
the two orthorhombic CuO$_2$ layers.
For two identical layers, we assume
$\langle a_{1{\bf -k}\downarrow}a_{1{\bf k}\uparrow}\rangle=
\langle a_{2{\bf -k}\downarrow}a_{2{\bf k}\uparrow}\rangle=
(\Delta_{\bf k}/2E_{\bf k})\tanh (E_{\bf k}/2T)$ by symmetry, where
$E_{\bf k}=[\xi_{\bf k}^2+\Delta_{\bf k}^2]^{1/2}$ is
the quasiparticle excitation spectrum with $\xi_{\bf k}$ the
particle band energy and $\Delta_{\bf k}$ the overall gap
self-consistently determined
by (\ref{eq:gap}). At $T=0$, Eq.~(\ref{eq:gap}) is reduced to
(layer index is redundant)

\begin{equation}
\Delta_{\bf k}=\Delta_0 f({\bf k})+\Delta_{\bf k}
{T_J({\bf k})\over 2E_{\bf k}},
\label{eq:gap.soln}
\end{equation}
where $\Delta_0 \equiv V\sum^{\omega_{c}}_{\bf k}\Delta_{\bf k}f({\bf k})/2E_{\bf k}$
with $\omega_c$ an appropriate BCS cutoff frequency.
At or near the FS ($|{\bf k}|\simeq k_F$),
$E_{\bf k}  \simeq |\Delta_{\bf k}|$
and (\ref{eq:gap.soln}) is further reduced to

\begin{equation}
\Delta_{\bf k}=\Delta_0 f({\bf k})+{T_J({\bf k})\over 2}{\rm sgn}[\Delta_{\bf k}].
\label{eq:gap1}
\end{equation}

Due to the term ${\rm sgn}[\Delta_{\bf k}]$,
it sets a {\em constraint} in (\ref{eq:gap1})
that physical solutions of $\Delta_{\bf k}$
arise only when $T_J({\bf k})$ preserves the symmetry of $f({\bf k})$
which in turn is determined by the nature of
the layer pairing interaction.
When the symmetry of $T_J({\bf k})$ is different from that of
$f({\bf k})$, the overall gap could exhibit a discontinuous phase on the FS.
This would imply a physically unstable state.
To fulfill the symmetry requirement, we thus argue that
$t_\perp({\bf k})\propto [\cos (2\phi)+s]^2$ and
$T_J({\bf k})\propto [\cos (2\phi)+s]^4$ assuming
$f({\bf k})=\cos (2\phi)+s$ for the present superconducting
orthorhombic bilayer.
More precisely, the overall gap function in (\ref{eq:gap1}) is given by

\begin{eqnarray}
\Delta_{\bf k}&=& \Delta_0\left[\cos(2\phi)+s\right]\nonumber\\
&+&{T_J \over 2}[\cos(2\phi)+s]^4
{\rm sgn}[\cos(2\phi)+s],
\label{eq:gap2}
\end{eqnarray}
where $T_J$ denotes the strength of the interlayer pair tunneling.
Eq.~(\ref{eq:gap2}) shows that not only the orthorhombicity $s$,
but also the pair tunneling
strength $T_J$ can be determined
through the investigation of the gap function.
It is worth emphasizing that when $s$ is small,
the $ab$-axis gap anisotropy could still be large,
as long as $T_J$ is significant.

The band structure calculation \cite{CSAS93,anderson94,xiang96}
and experimental evidences \cite{xiang01,feng01} usually assume
$t_\perp({\bf k})\propto( \cos k_x - \cos k_y)^2$
[or $\propto \cos^2(2\phi)$ in the continuum limit]
for high-$T_c$ cuprates in a square lattice of perfectly tetragonal symmetry.
This justifies our proposal $t_\perp({\bf k})\propto [\cos (2\phi)+s]^2$
in the tetragonal limit ($s=0$).

\begin{figure}
\mbox{
\epsfxsize=0.9\hsize
{\epsfbox{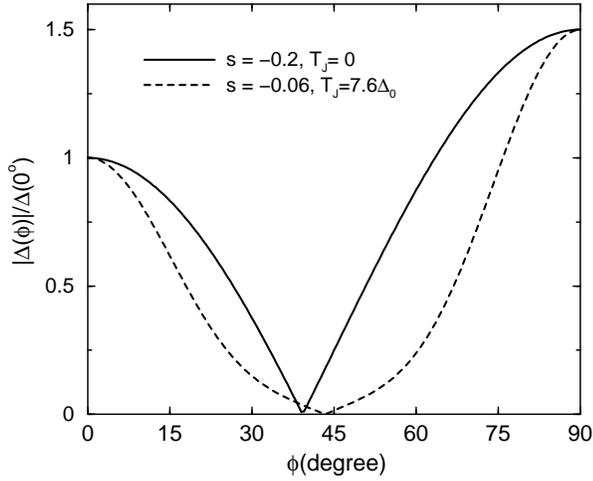}}}
\caption{The gap magnitude as a function of the azimuthal angle on the
FS for a superconducting orthorhombic bilayer
with interlayer pair tunneling [using
(\protect\ref{eq:gap2})].}
\label{fig1}
\end{figure}

Fig.~\ref{fig1} shows the gap magnitude as a function of
$\phi$ for two extreme cases: $s=-0.2, T_J=0$ and
$s=-0.06, T_J=7.6\Delta_0$. (In fact, $\Delta_0$ is a function of $T_J$ through
the self-consistent gap equation.)
These two cases all give $\Delta(90^{\rm o})/\Delta(0^{\rm o})=1.5$ in
regard to the ARPES result.
Several important features are noted for the anisotropic gap.
The nodal angle, deviated from the diagonals, is now determined by

\begin{eqnarray}
\phi_{\rm node}={1\over 2}\arccos(-s)\simeq {\pi\over 4}+{s\over 2} ~~
({\rm for~small} ~s)
\label{eq:phinode}
\end{eqnarray}
in the first quadrant of the FS.
Measurement of the nodal angle thus
reveals the value of $s$.
With the knowledge of $s$, one can then compare the
gap magnitude between the two antinodes

\begin{eqnarray}
{|\Delta(90^{\rm o})|\over |\Delta (0^{\rm o})|}=
{(1-s)[1+r(1-s)^3]\over (1+s)[1+r(1+s)^3]}~;~~~r\equiv {T_J\over 2\Delta_0},
\label{eq:ratio}
\end{eqnarray}
to obtain the value of $T_J$ (in unit of $\Delta_0$).
Alternatively one can also study the gap slope near the nodes

\begin{eqnarray}
\left|{\partial[\Delta(\phi)/\Delta(0^{\rm o})]\over
\partial\phi}\right|_{\phi\approx
\phi_{\rm node}}=2 {\sqrt{1-s^2}\over (1+s)[1+r(1+s)^3]},
\label{eq:slope}
\end{eqnarray}
which is 2 for a pure $d_{x^2-y^2}$-wave gap ($s=T_J=0$).

If the instrumental resolution is fine enough, ARPES would be
the best probe to explore $T_J$ and $s$ through the direct measurement
of gap anisotropy.
In the following, we study longitudinal ultrasonic attenuation
and inelastic neutron scattering spectra which are
also useful probes on these issues.

{\em Longitudinal Ultrasonic Attenuation} ---
Ultrasonic attenuation is a directional probe and thus
powerful to study the gap anisotropy. It has been successfully used to
study the order parameter symmetry in superconducting
heavy-fermion UPt$_3$ \cite{ellman96} and
high-$T_c$ analog Sr$_2$RuO$_4$ compounds \cite{lupien01}.
In clean YBCO where the scattering is in the ballistic limit
\cite{hosseini99}, the longitudinal ultrasound attenuation in the
superconducting state is proportional to \cite{Vekhter98-2,wu99}

\begin{eqnarray}
&&\alpha_s({\bf q},T)\propto \sum_{\bf k}
\left[-{\partial f(E_{\bf k})\over \partial E_{\bf k}}\right]{\xi_{\bf
k}^2\over E_{\bf k}}\times\nonumber\\ &&~
\delta\left(\xi_{\bf k} {\partial\xi_{\bf k}\over
\partial{\bf k}} \cdot {\bf q}+\Delta_{\bf k} {\partial\Delta_{\bf
k}\over \partial{\bf k}} \cdot {\bf q}\right),
\label{eq:alpha}
\end{eqnarray}
where ${\bf q}$ is the wavevector of the propagating phonon
and $f$ is the Fermi distribution function.
We shall calculate (\ref{eq:alpha}) for one single-layer with the input of the
interlayer pair tunneling gap in (\ref{eq:gap2}).
This is sufficient if no other coupling or
vertex correction is considered within the two layers.
Inspection of (\ref{eq:alpha}) shows that $\alpha_s$ is governed
by a delta function, weighted by the FS sum according to how
small the gap is. For an isotropic
$s$-wave superconductor, the second term in the delta function vanishes,
{\em i.e.}, only the portion of the FS perpendicular to ${\bf q}$
contributes. For unconventional superconductors such as YBCO, the
second term does contribute but is typically small
($\propto \Delta/\epsilon_F$ with $\Delta$ the maximum gap and $\epsilon_F$
the Fermi energy). By ignoring the second term in the delta function,
Eq.~(\ref{eq:alpha}) is simplified to

\begin{eqnarray}
{\alpha_s\over \alpha_n}= {\langle 2\delta({\bf k}_F\cdot {\bf q})
 f(|\Delta_{\bf k}|) \rangle\over \langle\delta({\bf k}_F\cdot
{\bf q}) \rangle}
\label{eq:ratio.limit}
\end{eqnarray}
between the superconducting and normal states,
where the angular bracket denotes the FS average.
Since only two points on the (two-dimensional) FS satisfy
${\bf k}_F\cdot {\bf q}=0$, Eq.~(\ref{eq:ratio.limit}) is simply solved to be
$\alpha_s/\alpha_n=2f(|\Delta(\theta)|)= 2/[\exp(|\Delta(\theta)|/T)+1]$,
where $\theta$ is related to $\phi_q$ (the angle of ${\bf q}$)
by $\theta-\phi_q=\pi/2$.

\begin{figure}
\mbox{
\epsfxsize=0.9\hsize
{\epsfbox{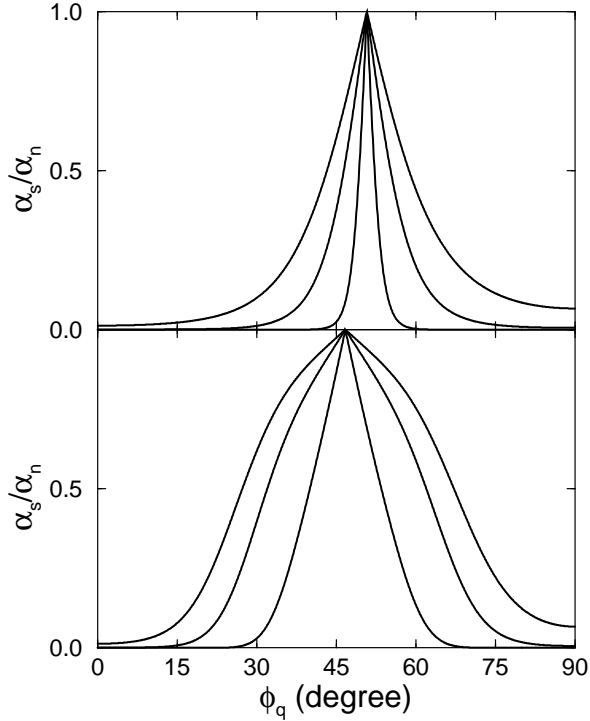}}}
\caption{The ratio of the longitudinal ultrasonic attenuation in superconducting
to normal state as a function of the angle of sound wavevector. Top (bottom) frame
corresponds to $s=-0.2$ and $T_J=0$ ($s=-0.06$ and $T_J=7.6\Delta_0$).
The curves from top to bottom are for $T/T_c=0.5,0.3$ and 0.1.}
\label{fig2}
\end{figure}

In Fig.~\ref{fig2}, $\alpha_s/\alpha_n$ vs. $\phi_q$ are shown
at different temperatures for those two cases
in Fig.~\ref{fig1}. We have assumed
$2\Delta(0^{\rm o})/T_c=3.52$ and the gap magnitude remains roughly the same
for $T\leq 0.5T_c$. [Of course, the exact $T$ dependence of gap can be worked out
through (\ref{eq:gap}), which is not done in this paper.]
The maximum $\alpha_s/\alpha_n$
appear at a specific angle $\phi_q$ at

\begin{eqnarray}
\phi_{\rm max}={\pi\over 2}-\phi_{\rm node},
\label{eq:phimax}
\end{eqnarray}
where the nodal angle $\phi_{\rm node}$ is given in (\ref{eq:phinode})
and we limit $0\leq \phi_{\rm max},\phi_{\rm node} \leq \pi/2$.
Since $\phi_{\rm node}$ depends on the value of $s$,
determination of $\phi_{\rm max}$ in $\alpha_s/\alpha_n$ thus
determines $s$. In addition, the ratio at $\phi_q=0$ to $\pi/2$,

\begin{eqnarray}
&&{\alpha_s/\alpha_n (\phi_q=0)\over \alpha_s/\alpha_n
(\phi_q=\pi/2)}= \nonumber\\
&&{1+\exp\left[{\Delta_0\over T}
\left((1+s)[1+r(1+s)^3]\right)\right] \over 1+\exp\left[{\Delta_0\over T}
\left((1-s)[1+r(1-s)^3]\right)\right]},
\label{eq:ratio.antinode}
\end{eqnarray}
allows one to extract the strength of $T_J$.
Furthermore, near the critical value, the slope

\begin{eqnarray}
\left|{\partial(\alpha_s/\alpha_n)\over \partial\phi_q}\right|_{\phi_q\approx
\phi_{\rm max}}={\Delta_0\over T} {\sqrt{1-s^2}\over (1+s)[1+r(1+s)^3]},
\label{eq:slope.alpha}
\end{eqnarray}
which gives one more way to look into $T_J$ and $s$,
as well as $\Delta_0$.

{\em Spin Excitation Spectra} ---
Recent inelastic neutron scattering (INS) experiments have
reported many interesting
results for high-$T_c$ cuprates regarding the commensurate and
incommensurate (IC) peaks at or near the antiferromagnetic (AF)
wavevector ${\bf Q}_{\rm AF}=(\pi,\pi)$ \cite{bourges00}.
For YBCO or BSCCO, the appearance of the commensurate peaks are likely due to
some kind of resonance, while the IC peaks are associated with
the dynamical local nesting effect of the band structures
\cite{brinckmann99,kao00,voo00}.

\begin{figure}
\mbox{
\epsfxsize=0.95\hsize{\epsfbox{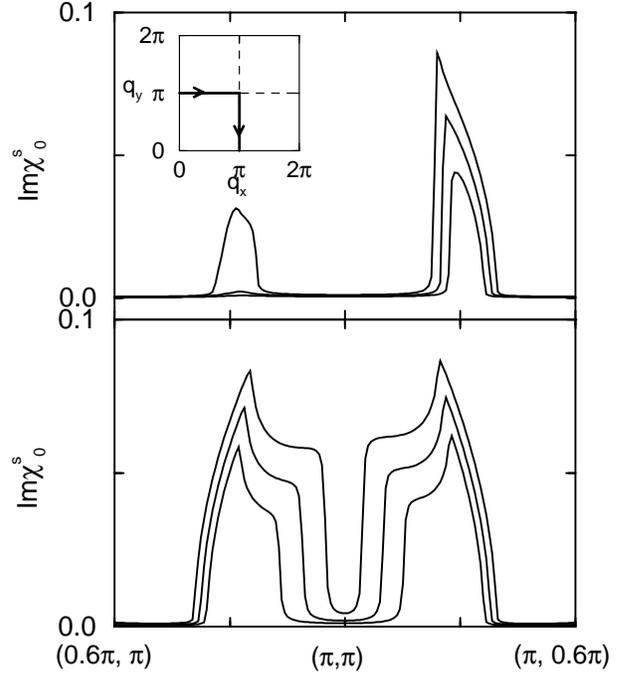}}}
\vspace{0.5cm}
\caption{Inelastic spin excitation spectra with the momentum scan
along the direction shown in the inset. Top (bottom) frame
corresponds to $s=-0.2$ and $T_J=0$ ($s=-0.06$ and $T_J=7.6\Delta_0$).
The curves from top to bottom are for $\omega/\Delta(0^{\rm o})
=1.0,0.8$ and 0.6.}
\label{fig3}
\end{figure}

Theoretically the INS spectra is proportional to
the imaginary part of the spin susceptibility and,
for simplicity, we study the irreducible BCS spin susceptibility
$\chi^s_0({\bf q},\omega)$ for one single-layer.
In Fig.~\ref{fig3}, ${\rm Im}\chi^s_{0}$ are calculated at $T=0$ for some fixed
frequency $\omega$ and momentum ${\bf q}$ scanned along the direction shown
in the inset. The primary interest is to see how the gap
(\ref{eq:gap2}) affects the INS spectra.
To study the IC peaks, a tight-binding band
$\xi_{\bf k}=-2t(\cos k_{x}+\cos k_{y}) -4t'\cos k_{x}\cos
k_{y}-\mu$ is inevitably used, where $t$ and $t^\prime$ are respectively
the nearest-neighbor and
next-nearest-neighbor hopping and $\mu$ is the
chemical potential. Typical values of
$t^\prime=-0.25t$ and $\mu=-0.65t$ are employed.
For the gap,  $\cos (2\phi)$ in (\ref{eq:gap2}) is replaced by
$(\cos k_{y}-\cos k_{x})/2$ in the present lattice case and we assume
the gap magnitude at $k_x$-axis, $\Delta(0^{\rm o})=0.3t$.
The same two cases in Figs.~\ref{fig1} and \ref{fig2} are studied.
For the $T_J=0$ (larger $s$) case,
the nodes are shifted away from the diagonals and consequently
the nesting effect is highly anisotropic:
the IC peak parallel to $q_y$ axis (along the ${\bf q}$ scan route chosen)
is more intense than the one parallel to $q_x$ axis.
At lower frequency, the latter could
completely disappear before the nesting effect can set in.
For the strong $T_J$ (small $s$) case, in contrast,
the nodes remain closer to the diagonals and, as a result,
the nesting effect is roughly isotropic. The IC peaks are nearly
symmetric along the ${\bf q}$ route, regardless the change of frequency.
These are the key features which can
be used to distinguish the strong and weak $T_J$
(or small and large $s$) cases.

When the commensurate peaks are also considered,
the Random phase approximation (RPA) corrected spin susceptibility
$\chi^s({\bf q},\omega)=\chi^s_0({\bf q},\omega)/[1-V({\bf q})
\chi^s_0({\bf q},\omega)]$ is often studied, where
$V({\bf q})$ is usually modeled by an ``AF'' interaction
$V({\bf q})=-J(\cos q_x+\cos q_y)/2$ ($J>0$).
Since here we are only interested in the IC peaks which appear
beyond the AF resonant regime, it's adequate to study $\chi^s_0$.
The line shape in ${\rm Im}\chi^s({\bf q},\omega)$ will be very
similar to those in $\chi^s_0({\bf q},\omega)$,
apart from different intensity.

Mook {\em et al.} \cite{mook00} have recently reported the 1D
nature for the IC peaks in underdoped YBa$_2$Cu$_3$O$_{6.6}$.
The IC peak intensity at
$(\pi-\delta, \pi)$ is found to be stronger than that at
($\pi,\pi-\delta)$.  While the results are referred to
favor the formation of the dynamical stripes for that particular doping
\cite{mook00}, an alternative explanation to these data is
to take into account the in-plane gap anisotropy associated with a
homogeneous Fermi liquid \cite{voo01}.
The data of Mook {\em et al.} seems indicating an intermediate value
of $T_J$ (and an intermediate and positive $s$) on YBa$_2$Cu$_3$O$_{6.6}$,
when comparing to the results in Fig.~\ref{fig3}.
It is important to have INS performed at different frequencies
in order to clarify the gap anisotropy.

In summary, we propose that the long-thought interlayer pair tunneling
effect can be explored through the detailed studies of the
gap anisotropy in YBa$_2$Cu$_3$O$_{7-\delta}$.
Considering an orthorhombic superconducting
bilayer with interlayer pair tunneling, one can {\em simultaneously} determine
the in-plane orthorhombicity and the strength of the pair tunneling.
For probable probes,
we study longitudinal ultrasonic attenuation and
inelastic neutron scattering.

Finally, we comment on the effect of pair tunneling on the collective modes.
For a superconducting
bilayer coupled by the pair tunneling, apart from the gap renormalization,
the system will exhibit the characteristic in-phase and
out-of-phase phase modes associated with the order parameters on the
two layers.
The in-phase Anderson-Bogoliubov phase mode has a phonon dispersion,
$\omega^2=v_F^2q^2/2$, which is lifted
to plasmon mode when long-range Coulomb potential is included.
These in-phase modes, which are intrinsic characteristics of a superconductor,
are independent of $T_J$.
In contrast, the out-of-phase phase mode
has an optical phonon dispersion $\omega^2=\omega_0^2+v_F^2q^2/2$, where
$\omega_0$ depends on the relative size of planar pair interaction $V$ and
$T_J$ \cite{wu95}. It is also interesting
to study $T_J$ through the search of the out-of-phase phase modes,
such as those recently observed in SmLa$_{0.8}$Sr$_{0.2}$CuO$_{4-\delta}$
with two different Josephson couplings \cite{dulic01}.

The author thanks Prof. I.F. Herbut and physics department
at Simon Fraser University for their hospitality and
Prof. J.P. Carbotte, Prof. A. Griffin, and Dr. K.-K. Voo for useful discussions.
Financial support from NSC of Taiwan under the grant No. 89-2112-M-003-027
is acknowledged.

\end{document}